# Sub-megahertz nucleation of plasmonic vapor microbubbles by asymmetric collapse

Fulong Wang[1], Huimin Wang[2], Binglin Zeng[1], Chenliang Xia[1], Lihua Dong[1], Lijun Yang[3,4], and Yuliang Wang[*,1,4]

[1]*Robotics Institute, School of Mechanical Engineering and Automation, Beihang University, Beijing 100191, People's Republic of China*

[2]*National Center for Materials Service Safety, University of Science and Technology Beijing, Beijing 100083, People's Republic of China*

[3]*School of Astronautics, Beihang University, Beijing 100191, People's Republic of China*

[4]*Ningbo Institute of Technology, Beihang University, Ningbo 315832, People's Republic of China*

[*]E-mail: wangyuliang@buaa.edu.cn

**Abstract**

Laser triggered and photothermally induced vapor bubbles have emerged as promising approaches to facilitate optomechanical energy conversion for numerous relevant applications in micro/nanofluidics. Here we report the observation of a sub-megahertz spontaneous nucleation of explosive plasmonic bubbles, triggered by a continuous wave laser. The periodic nucleation is found to be a result of the competition of Kelvin impulsive forces and thermal Marangoni forces applied on residual bubbles after collapse. The former originates from asymmetric bubble collapse, resulting in the directed locomotion of residual bubbles away from the laser spot. The latter arises in a laser irradiation induced heat affected zone (HAZ). When the Kelvin impulses dominates, residual bubbles move out of the HAZ and the periodic bubble nucleation occurs, with terminated subsequent steadily growing phases. We experimentally and numerically study the dependence of the nucleation frequency $f$ on laser power and laser spot size. Moreover, we show that strong fluid flows over 10 mm/s in a millimeter range is steadily achievable by the periodically nucleated bubbles. Overall, our observation highlights the opportunities of remotely realizing strong localized flows, paving a way to achieve efficient micro/nanofluidic operations.

**Introduction**

Due to the enhanced plasmonic effect, laser irradiated and water immersed noble metal nanoparticles can rapidly produce huge amounts of heat and trigger the nucleation of so called plasmonic bubbles (*1-5*). Compared with direct bubble cavitation induced by high energy pulse lasers

(*6*), the applied light absorbers significantly increase photothermal conversion efficiency. As a result, plasmonic bubbles can be readily nucleated with an ordinary low power continuous wave (CW) laser (*7*). These bubbles have been emerged in numerous innovative applications, including cancer therapy (*8, 9*), micromanipulation (*10*), nanoscale propulsion (*2-4*), microfabrication (*11*), and ultrasensitive detection (*12*). The nucleation of plasmonic bubbles involves rich physicochemical hydrodynamics (*13, 14*). Exploring new phenomena during nucleation and growth of these bubbles helps to expand their applications.

Now it is clear that the nucleation of plasmonic microbubbles on a laser irradiated gold nanoparticle (GNP) decorated sample surface experiences four sequential life phases (*5, 15*): an explosively growing and rapidly collapsing giant vapor bubble (phase 1, hereafter referred to as initial plasmonic bubble: IPB) (*15, 16*), an oscillating bubble (phase 2) (*17, 18*), and two steadily growing phases 3 and 4 dominated by vaporization and gas diffusion, respectively (*5*). The life time of IPBs is around 10 $\mu$s with the maximum growth rate up to 20 m/s, which is at least three orders of magnitude faster than their steadily growing counterparts (*15, 19, 20*).

The ultrashort life time and extreme fast expansion speed of IPBs indicate that strong localized flows with hundreds of kilohertz modulation frequency can be achieved remotely. This is astonishing in low Reynolds number and viscosity dominated micro/nanofluidics (*21, 22*), where localized strong flows are highly demanded to efficiently implement various operations, like mixing, pumping, sorting, and propulsion (*23-28*). Clearly, the excellent hydrodynamic features of IPBs imply their great potentials in micro/nanofluidics related applications. However, as mentioned earlier, IPBs are inevitably succeeded by the subsequent three life phases. A question remains as how to achieve sustained periodic nucleation of IPBs with eliminated subsequent life phases of plasmonic bubbles.

In this study, we report an observation of spontaneously triggered periodic nucleation of IPBs with a frequency *f* over 200 kHz nearby a rigid boundary. We will show how the interplay of the rigid boundary induced impulsive forces and thermal Marangoni forces determines the fate of residual bubbles after IPB collapse, which in turn governs periodic bubble nucleation. Moreover, the dependence of nucleation frequency *f* on the control parameters, namely laser power and laser spot size will be quantitatively studied. Our finding not only shows rich hydrodynamics along plasmonic bubble nucleation, but also paves a way for efficient micro/nanofludic operations mentioned above.

## Results and discussion

### *Observation of periodic nucleation of initial plasmonic bubbles*

Two experiments were first conducted for comparison on a GNP decorated sample surface (**Fig. S1A**) with a 532 nm CW laser at a laser power $P_l$ = 135 mW through a 10× objective lens. The applied frame rate is 900 kfps (kilo frames per second), corresponding to a 1/900,000 = 1.11 $\mu$s temporal resolution. In the first experiment, laser beam was directly shined on the sample surface. In the second experiment, laser beam was shined on the sample surface near a polydimethylsiloxane (PDMS) vertical side wall (**Fig. S1B**). All experiments were conducted in a home built optical setup (**Fig. S1C**).

In the first experiment, upon laser irradiation, radius $R$ of a nucleated plasmonic bubble as a function of time $t$ in 5 s is depicted in **Fig. 1A**. The snapshots of bottom view images of the plasmonic bubble captured at five instants of ① to ⑤ are shown in **Fig. 1B**. After a 28.8 $\mu$s delay from the beginning of the laser irradiation, an IPB was nucleated, rapidly grew to its maximum size of $R$ = 21 $\mu$m, and quickly collapsed in 4.5 $\mu$s, as shown in **Fig. 1B**①, ②, and ③, respectively. After that, a very tiny residual bubble remained at the solid-liquid interface. It then acted as a nucleus for a subsequent steadily growing plasmonic bubble. The bubble gradually grew to 83.9 $\mu$m in 5 s (**Fig. 1B** ④ and ⑤). This observation is consistent with that reported previously (*5, 15*).

In the second experiment, the radius $R$ of the nucleated plasmonic bubbles as a function of $t$ in 5 ms is depicted in **Fig. 1C**. Remarkably, we find that IPBs can repeatedly nucleate, grow, and collapse at a very high frequency. **Fig. 1D** is the enlarged figure for an area selected by a dashed box in **Fig. 1C**. This is totally different from that shown in **Fig. 1A**. From the $R - t$ curve, the cycle period $T$ of the periodic bubble nucleation can be measured. For details, please refer to **Fig. S2** in Supplementary Information. The measured frequency $f$ as a function of time $t$ in the first 0.5 ms is shown in **Fig. 1E**. At the beginning, $f$ is about 32.3 kHz. It rapidly increases and then stabilizes at 100 kHz in 0.2 ms.

One may wonder what is the reason for the eliminated steadily growing plasmonic bubbles in the second experiment. Bottom view images of an IPB captured at five instants of ① to ⑤ in **Fig. 1D** are depicted in **Fig. 1F**. The IPB grows and collapses within 6.6 $\mu$s. During the process, the IPB is displaced from the laser spot to the side wall, leaving a residual bubble next to the vertical wall. This is different from that happened in the first experiment, where the residual bubble stays on the same location of bubble nucleation (**Fig. 1B**③). This clearly indicates that the side wall results in the asymmetric bubble collapse and hence the lateral displacement of the residual bubble, resulting in periodic nucleation of IPBs.

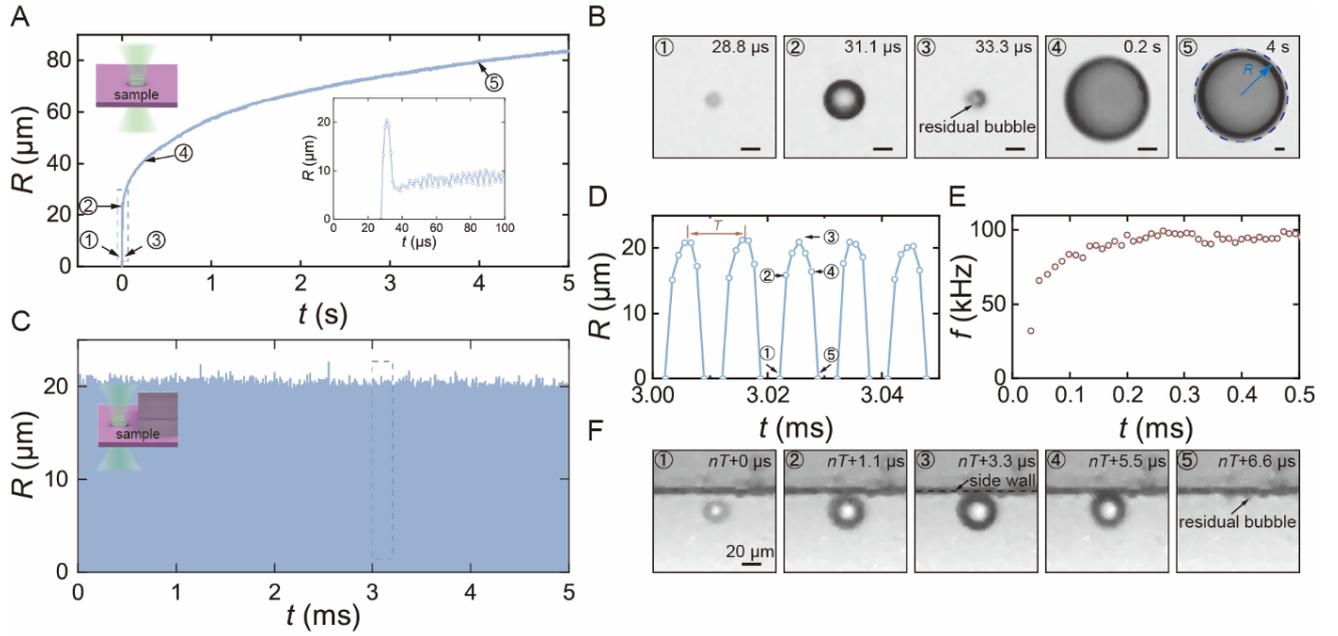

**Fig. 1** Dynamics of the nucleated plasmonic bubbles with and without a vertical side wall on a GNP decorated sample surface ($P_l$ = 135 mW). (A) Radius $R$ of plasmonic bubbles on the sample surface without a side wall. An initial plasmonic bubble (IPB) first quickly nucleated and then collapsed, followed by a steadily growing ordinary plasmonic bubble. The left inset diagram illustrates the irradiation of laser beam on the sample surface. The right inset graph is the enlarged plot of the area selected by a dashed box in the first 100 μs upon laser irradiation. (B) Bottom view snapshots of an IPB (①-③) and the subsequent steadily growing plasmonic bubble (④ and ⑤) captured at five different instants of ①-⑤ in A (all scale bars: 20 μm). (C) Radius $R$ of five hundred of repeatedly nucleated IPBs as a function of time $t$ in 5 ms when the laser spot is near a vertical PDMS side wall, corresponding to a $f$ = 100 kHz nucleation frequency. The inset diagram illustrates the relationship between the laser spot and the vertical side wall. (D) $R$ - $t$ dependence in five selected cycles from figure C. The cycle period $T$ is defined. (E) The nucleation frequency $f$ as a function of $t$ in the first 0.5 ms from figure C. (F) Bottom view snapshots captured at five selected instants of ①-⑤ for an IPB in figure D. At the end of the collapse, the bubble moves towards the side wall.

*Mechanism of the periodic bubble nucleation*

During experiments, we found that the lateral distance $L_{las}$ of the laser spot to the side wall is a key factor in determining the periodic nucleation of IPBs. A comparison of three IPBs with a similar $R_{max}$ ≈ 42 μm and different $L_{las}$ is shown in **Fig. 2A**. For a bubble with a relatively smaller value of $L_{las}$ = 60 μm, its residual bubble is highly displaced towards the side wall (**Fig. 2A(I)**). When $L_{las}$ is further increased to 110 μm, the effect of the side wall vanishes and the residual bubble remains in the laser spot area (**Fig. 2A(II)**). This is very similar to that happened on a free interface without the side wall (**Fig. 2A(III)**). In the latter two cases, the periodic bubble nucleation cannot take place.

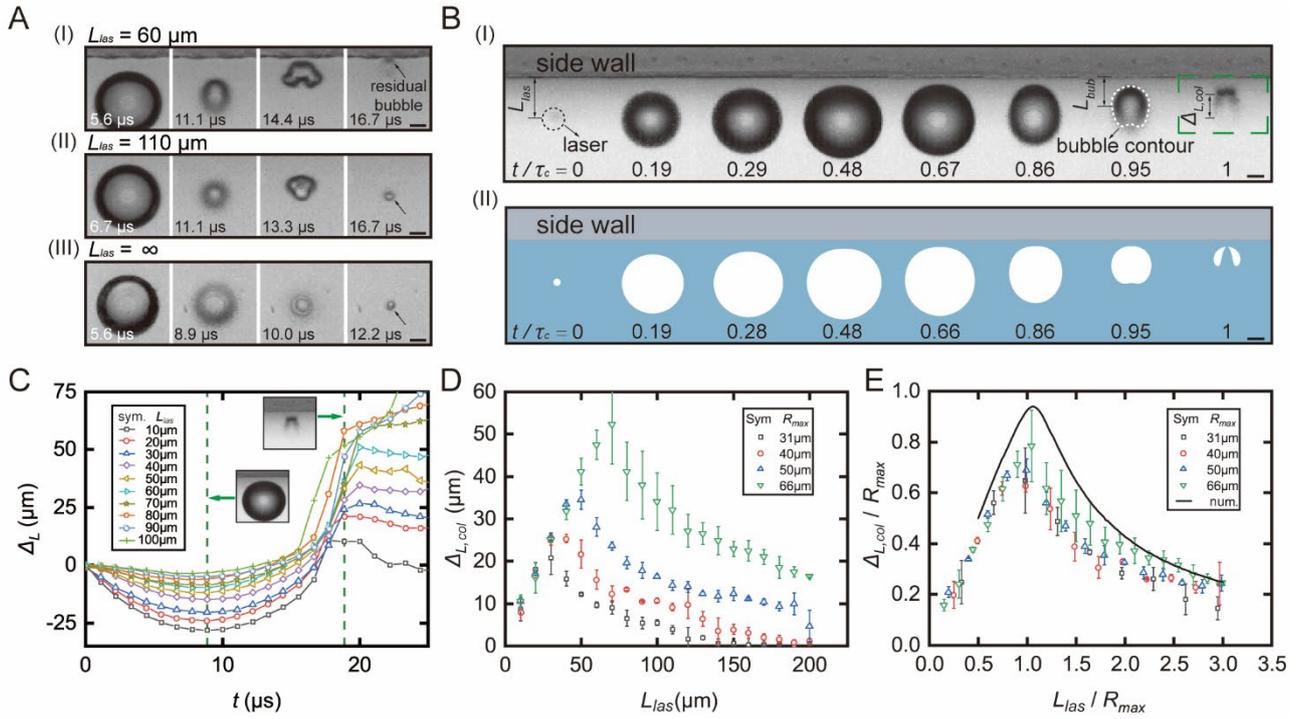

**Fig. 2** Characterization of bubble motion near a vertical side wall. (A) Comparison of bubble collapse process for three IPBs with different $L_{las}$. (B, I) Temporal evolution of an IPB near a side wall along its entire life cycle observed from bottom view. The bubble collapses at $t/\tau_c = 1$, leaving a tiny residual bubble. $L_{las}$ and $L_{bub}$ are lateral distance of the centers of the laser spot and the IPB to the side wall, while $\Delta_{L,col}$ is the lateral distance of the residual bubble to the center of the laser spot, namely the maximum lateral displacement of the IPB (scale bar: 40 $\mu$m, $R_{max} \approx 90$ $\mu$m and $L_{las} = 100$ $\mu$m). (B, II) Numerically simulated temporal evolution of a vapor bubble near a side wall. The normalized initial size of the vapor bubble is 0.1 and the ratio $L_{las} / R_{max} = 10/9$ (scale bar: 40 $\mu$m). (C) Displacement $\Delta_L$ as a function of $t$ for IPBs of $R_{max} \approx 66$ $\mu$m with different values of $L_{las}$. (D) $\Delta_{L,col}$ as a function of $L_{las}$ for bubbles with different $R_{max}$ of 31, 40, 50, and 66 $\mu$m. $\Delta_{L,col}$ first increases and then decreases with $L_{las}$. (E) $\Delta_{L,col}/R_{max}$ as a function of $L_{las}/R_{max}$ for IPBs with different $R_{max}$. The solid line is the result from the numerical simulation, which shows a good agreement with the experimental results.

To get more insights, we systematically adjusted $L_{las}$ from 10 to 200 $\mu$m with a step size of 10 $\mu$m to investigate how bubble's displacement $\Delta_L = L_{las} - L_{bub}$ changes accordingly, where $L_{bub}$ is bubble's lateral distance to the side wall. **Fig. 2B(I)** shows sequentially captured bottom view images of an IPB with $L_{las} = 100$ $\mu$m along its life cycle. The life cycle $\tau_c$ of an IPB starts from the beginning of the bubble nucleation and ends when the bubble first collapses to a minimum size. Therefore, it contains two periods: growth and collapse. Note here that the time in **Fig. 2B** is normalized as $t/\tau_c$. Within its life time, the bubble rapidly grows and reaches its maximum size at about $t/\tau_c = 0.48$. Due to the presence of the side wall, the bubble exhibits asymmetric shape during both growth and collapse periods. At the end of collapse period ($t/\tau_c$: 0.76→1), the bubble quickly moves towards the side wall and a minimum value of $L_{bub}$ is achieved. This corresponds to the maximum bubble displacement, which is defined as $\Delta_{L,col}$. After the collapse period, a residual bubble will appear at the location where the bubble collapses. The residual bubble contains certain amount of non-condensable air and can last for a relatively longer period of time. The stored energy in IPBs is violently released during aforementioned bubble growth and collapse periods. As a result, the subsequent process only possesses a small portion of energy (*29*), which is not the focus of this work.

To have a better understanding of the asymmetric collapse of IPBs, a two dimensional (2D) simulation was conducted by using a boundary intergrade method (BIM) (*30, 31*). For simplicity, it ignores the supporting substrate and focuses on the impact of the side wall on the dynamics of the vapor bubbles. One of the results which exhibits close values of $R_{max}$ and $L_{las}$ to the case shown in **Fig. 2B(I)** is depicted in **Fig. 2B(II)**. It clearly shows that the numerically simulated bubble exhibits the same growth and collapse dynamics with that in the experiment. The rapid motion of a collapsing bubble towards the side wall is due to the so-called Kelvin impulse (*30, 32*). The nearby side wall leads to the asymmetric pressure and velocity fields in water around the bubble. This results in a net liquid pressure and hence a translational momentum of the bubble. As a result, the IPB is significantly displaced from the laser spot during the collapse period. This refreshes the surrounding fluid field and makes it ready for the nucleation of the following IPBs. This is the reason for the repeated and periodic nucleation of IPBs.

Experiments were further conducted to systematically investigate how $\Delta_L$ changes with $L_{las}$. **Fig. 2C** first depicts how $\Delta_L$ changes with time for bubbles with $R_{max} = 66$ $\mu$m. In spite of different $L_{las}$, all IPBs first moved away from the side wall during the growth period, exhibiting a negative value of $\Delta_L$. After that, the bubbles rapidly moved towards the side wall, resulting in the increased $\Delta_L$. At the end of bubble collapse ($t \approx 18.8$ $\mu s$), the bubbles reached their maximum displacement $\Delta_{L,col}$ in their entire life cycles. By adjusting $P_l$, bubbles with different $R_{max}$ were obtained (*15*). The value $\Delta_{L,col}$ as a function of $L_{las}$ for bubbles with different $R_{max}$ is shown in **Fig. 2D**. All bubbles exhibit a similar trend that $\Delta_{L,col}$ first increases with $L_{las}$, reaching a maximum value and then decreases. The observed $\Delta_{L,col}$ - $L_{las}$ dependence is similar to that reported somewhere else (*31, 33-36*).

Interestingly, by normalizing $\Delta_{L,col}$ and $L_{las}$ with respect to $R_{max}$, we find that the normalized $\Delta_{L,col}$ - $L_{las}$ curves collapse on each other for bubbles with different $R_{max}$ (**Fig. 2E**). $\Delta_{L,col}/R_{max}$ reaches its maximum value when $L_{las}$ equals to $R_{max}$. This relationship was furthered numerically investigated with the BIM method (black solid curve in **Fig. 2E**). It clearly shows that the experimentally obtained $\Delta_{L,col}/R_{max}$ - $L_{las}/R_{max}$ dependence agrees well with numerical result, except that the numerical curve is slightly higher than the experimental results. This is believed to be due to the presence of drag force in between IPBs and the substrate, which is absent in the numerical simulation.

Now it is clear that the side wall changes the dynamics of bubble growth and collapse, resulting in a lateral displacement $\Delta_{L,col}$ of the residual bubbles. However, it remains unclear about the required $\Delta_{L,col}$ to trigger the periodic nucleation. To answer this question, probability maps were constructed. One of the constructed maps is shown in **Fig. 3A**. To construct this map, the value $L_{las}$ for IPBs with $R_{max} = 50$ $\mu$m was adjusted from 20 to 150 $\mu$m with a step size of 10 $\mu$m. At each distance $L_{las}$, experiments were run for ten times. In each time of running, the laser irradiation lasted for 50 ms and the duration of the periodic nucleation lasts was recorded. After that, the probability that the periodic nucleation lasts for a certain period of time was calculated and the probability map was then constructed. The map shows that the periodic nucleation can be successfully implemented when $L_{las} < 90$ $\mu$m. When $L_{las}$ is over 120 $\mu$m, the periodic nucleation cannot take place any more. Within the range of 90 $\mu$m $< L_{las} < 120$ $\mu$m, the duration of periodic nucleation rapidly decreases. This range is taken as the transition region (TR).

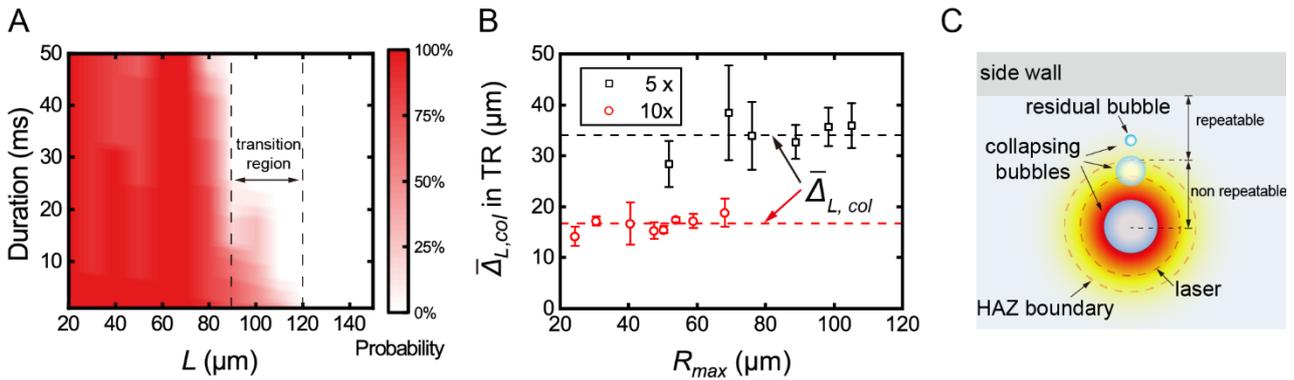

**Fig. 3** Correlation of residual bubble position with the periodic bubble nucleation. (A) A probability map constructed by triggering periodic nucleation of IPBs ($R_{max} = 50$ $\mu$m) with changing $L_{las}$ from 20 to 150 $\mu$m. (B) The average value $\overline{\Delta}_{L,col}$ in the transition region (TR) for bubbles with different $R_{max}$ and obtained with the 5× and 10× objective lenses. $\overline{\Delta}_{L,col}$ remains almost constant for each objective, regardless of changing $R_{max}$. (C) A schematic diagram illustrating the underlying mechanism of the periodic nucleation of IPBs. After bubble collapse, the residual bubble must move out of a heated affected zone (HAZ) to guarantee the periodic nucleation.

For IPBs with different $R_{max}$ and obtained with 10× and 5× objective lenses, probability maps were constructed. From individual probability maps, we extracted $\Delta_{L,col}$ of ten bubbles at the center of their transition regions (here $L_{las} = 100$ $\mu$m in **Fig. 3A**) and calculated their average value $\overline{\Delta}_{L,col}$. The value $\overline{\Delta}_{L,col}$ for bubbles with different $R_{max}$ is shown in **Fig. 3B**. It clearly shows that $\overline{\Delta}_{L,col}$ remains almost constant at 16.7 and 34.0 $\mu$m for the 10× and 5× objective lenses, respectively, regardless of different $R_{max}$. In the transition region, the ratio $L_{las}/R_{max}$ is in between 1.5 and 2.5. From **Fig. 2E** one can see that $\Delta_{L,col}$ decreases with increasing $L_{las}$ in this range. Therefore, $\bar{\Delta}_{L,col}$ can be taken as a critical lateral displacement of IPBs. If the lateral displacement of a residual bubble is below this critical value, it is hard to trigger the periodic bubble nucleation.

One may wonder what essentially determines these specific critical values of $\bar{\Delta}_{L,col}$ for periodic bubble nucleation. Upon laser irradiation, the GNP decorated sample surface rapidly produces a huge amount of heat, leading to a locally elevated and Gaussian-like temperature field (*13, 37, 38*). This results in a strong temperature gradient along the radial direction of the laser spot. The temperature gradient further induces a highly asymmetrical surface tension over bubble surface and causes a strong thermal Marangoni force, pushing the bubble back to the hottest position, the center of the laser spot. Therefore, the critical value of $\bar{\Delta}_{L,col}$ actually defines the boundary of a heat affected zone (HAZ), as illustrated in **Fig. 3C**. When residual bubbles stay in HAZ, they will be act as nuclei for the nucleation and growth of the following steadily growing plasmonic bubbles. The periodic bubble nucleation cannot take place. Therefore, the residual bubbles must move out of the HAZ to trigger the periodic nucleation. Due to thermal diffusion, the size of the HAZ is larger than that of the laser spots. That is why the measured laser spot radii $R_l$ of 11.5 and 28.0 $\mu m$ are smaller than corresponding $\bar{\Delta}_{L,col}$ of 16.7 and 34.0 $\mu m$ for the 10× and 5× objective lenses, respectively. For details of laser spot radius determination, please refer to **Fig. S3** in the supplementary information.

### *Controllability of the periodic bubble nucleation*

In micro/nanofluidics related applications, the frequency of periodic impulses is important. Here we investigate the controllability of these periodic nucleated bubbles, especially for the frequency *f*. One entire cycle *T* of a periodically nucleated IPB includes two periods, a delay $\tau_d$ and the bubble life time $\tau_c$. As revealed previously, a higher laser power $P_l$ leads to shorter $\tau_d$ and $\tau_c$ (*15*), which in turn results in a higher frequency *f*. Moreover, the size of laser spots also plays a key role in the nucleation dynamics of IPBs. In this study, laser spot size was adjusted by changing the magnification *m* of the objective lenses. Therefore, *f* was tuned by changing *m* and $P_l$. **Fig. 4A** and **4B** show how $\tau_d$ and $R_{max}$ change with applied $P_l$ for IPBs obtained with three different objectives of 5×, 10×, and 20×. For each data point, experiments were run for five times. The mean value and the standard deviation of data are obtained. Readers should be noted that the data points in **Fig. 4A** and **4B** are from the first triggered IPBs. For a specific objective, $\tau_d$ rapidly decreases with increasing $P_l$, leading to a quickly decreased $R_{max}$. Across different objectives, bubbles obtained with a higher magnification objective exhibit a much shorter $\tau_d$ and lower $R_{max}$. Overall, it shows that the delay time $\tau_d$ can be readily tuned from 3 to 1500 $\mu$s, corresponding to a large $R_{max}$ distribution from 6.8 to 106.1 $\mu m$.

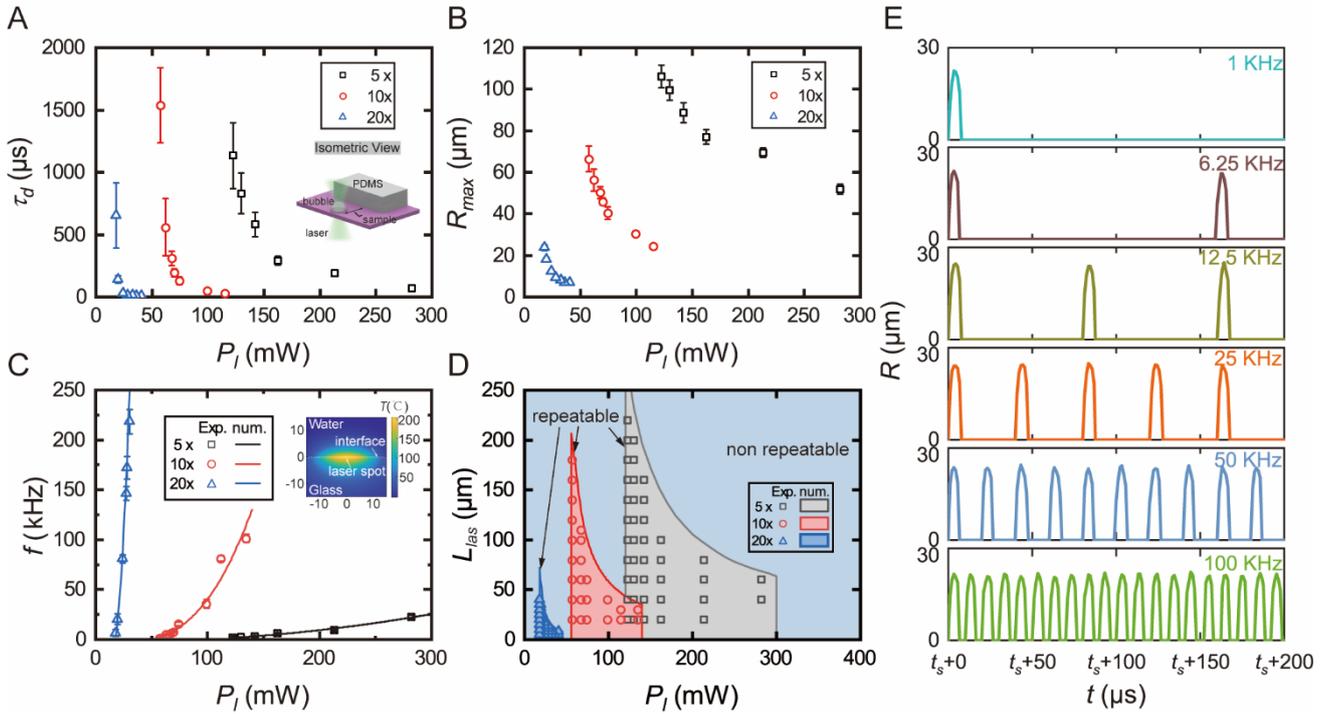

**Fig. 4** Controllability of the periodic nucleation of IPBs near a side wall. (A) and (B) are measured delay time $\tau_d$ and bubble maximum radius $R_{max}$ as functions of $P_l$, respectively. Both $\tau_d$ and $R_{max}$ decrease with $P_l$. (C) Frequency $f$ of the periodic nucleation as a function of $P_l$ for IPBs obtained with objective lenses of different magnification $m$. The symbols are experimentally measured data points, while the solid curves are from numerical calculation. The frequency $f$ can reach 218.4 kHz at $P_l$ = 30.4 mW and $m$ = 20×. The inset figure is a numerically constructed temperature field. (D) Determined regimes of the periodic bubble nucleation in a $L_{las}$ - $P_l$ two-dimensional space. The symbols represent experimental data and the solid curves are numerically determined boundaries of the periodic nucleation regimes. (E) Spontaneous nucleation frequency $f$ = 100 kHz ($P_l$ = 135 mW and $m$ = 10×) and modulated frequencies of 50, 25, 12.5, 6.25, and 1 kHz achieved through laser on/off modulation.

The different combinations of objective magnification $m$ and laser power $P_l$ lead to different nucleation dynamics of IPBs. This in turn leads to adjustable $f$, as shown in **Fig. 4C**. In the figure, each data point was average value from five running. For each objective lens, the achieved $f$ increases with $P_l$. We realized spontaneous periodic nucleation with different frequencies of 80, 39, 18, 9, and 5.4 kHz with different $P_l$ of 112.3, 99.0, 74.3, 67.6, and 62.5 mW, respectively, by using the 10× objective lens. A surprising controllability was well achieved. Remarkably, a maximum value of $f$ = 218.4 kHz is achieved with the 20× objective at $P_l$ = 30.4 mW. This sub-megahertz frequency is at least two orders of magnitude faster than what has been achieved so far with a continuous wave laser (25, 26).

Please note here that the corresponding cycle period $T = 1/f$ in **Fig. 4C** is shorter than that derived from **Fig. 4A** and **4B**. This is due to the sustained energy input from the laser, which inevitably warms up the surrounding water in the vicinity of the laser spots, leading to decreased $\tau_d$ and hence increased $f$. The time scale that the surrounding area reach a steady temperature field can be given as $\tau_{diff} \approx R_l^2/\pi\kappa$, where $\kappa$ is the diffusivity of water. By taking the 10× objective lens as an example, the required time is about 250 $\mu$s. This means that $f$ will rapidly increase at the beginning of the periodic nucleation and then reach a steady value. This is consistent with the data shown in **Fig. 1E**. In this study, we numerically determined the temperature elevation $\Delta T_E$ of the surrounding water by using the experimentally measured cycle $T$ during the steady stage of the periodic bubble nucleation. After that, the delay $\tau_d$ for different combinations of $P_l$ and $m$ can be obtained with numerically constructed temperature fields around laser spots (see one example in the inset figure in **Fig. 4C**). Since the life time $\tau_c$ of IPBs can take up to 50% of $T$ when $f$ is over 100 kHz, we also considered $\tau_c$ of IPBs by using the numerically determined $R_{max}$ (*39*) and a modified Rayleigh collapse time equation (*40*). Based on the numerically determined $\tau_d$ and $\tau_c$, $f$ at a specific combination of $P_l$ and $m$ can be numerically determined (solid curves in **Fig. 4C**). One can see that they show great agreement with the experimental measurement. For detailed process of determination of $\Delta T_E$, bubble life time $\tau_c$, and the numerical calculation of $f$, please refer to **Fig. S4A, S4B, S4C** in the supplementary information.

Now it is clear that both $L_{las}$ and $P_l$ are key parameters in controlling the periodic bubble nucleation. Here we systematically conducted experiments to determine the available regimes to trigger the periodic bubble nucleation in a $P_l$ - $L_{las}$ two-dimensional parameter space for 5×, 10×, and 20× objective lenses. The results are shown in **Fig. 4D**. In the figure, each marker represents a successful periodic nucleation of IPBs under a specific combination of ($P_l$, $L_{las}$). The regimes of successful periodic bubble nucleation were also determined numerically. For each objective lens, the bubble size $R_{max}$ can be numerically determined under a specific value of $P_l$ (**Fig. S4C**)(*39*). With the obtained $R_{max}$ and the criterion determined by **Fig. 3B**, the boundaries of the regimes of periodic bubble nucleation in the $P_l$ - $L_{las}$ parameter space were numerically determined (solid curves in **Fig. 4D**). One can see that the experimental data points are well enclosed by the numerically determined boundary curves. For each objective lens, bubble size decreases with increasing $P_l$, leading to a reduced available range of $L_{las}$ and increased nucleation frequency $f$.

Here readers should be noted that the nucleation frequency $f$ cannot be continuously increased by blindly increasing $P_l$. On one hand, the size of the HAZ determines a minimum value of $R_{max}$ to ensure periodic nucleation. Otherwise, the residual bubbles cannot move out of the HAZ. This determines one maximum laser power $P_{l, max1}$. On the other hand, the sustained laser energy input to the laser spot area cannot be dissipated in time with a high $P_l$. This will cause a fast vaporization prior to the successful escape of the residual bubbles from the HAZ. This determines another maximum laser power $P_{l, max2}$.

In practice, the smaller one of $P_{l,\,max1}$ and $P_{l,\,max2}$ determines the right boundaries of individual repeatable bubble nucleation regimes in **Fig. 4D**. Moreover, as mentioned earlier, the life time of a micro-sized vapor bubble is a few microseconds (*41*). This implies that the maximum $f$ should be several hundreds of kilohertz.

In the regimes of repeatable bubble nucleation in **Fig. 4D**, a certain combination of ($L_{las}$, $P_l$, $m$) corresponds to a specific $f$, which actually corresponds to the maximum $f$ one can achieve under that combination. By applying laser on/off modulation, any frequency under this value can be achieved. For $f = 100$ kHz obtained under the combination of $L_{las} = 15$ $\mu$m, $P_l = 135$ mW, and $m = 10\times$, different bubble nucleation frequencies of 50, 25, 12.5, 6.25, and 1 kHz were readily achieved through laser on/off modulation at corresponding frequencies (**Fig. 4E**).

*Strong localized flows induced by the periodic bubble nucleation*

Our finding provides an efficient approach to achieve strong and sustained localized fluid flows with a CW laser in micro/nanofluidics. To demonstrate this, two kinds of configurations were applied. The particle image velocimetry (PIV) was used to measure the fluid flow by tracking polystyrene (PS) particles during the periodic bubble nucleation. In the first one, the periodic nucleation was triggered next to a vertical side wall, which simulates a scenario of microfluidics in microchannels of bio-MEMS. Two symmetrically distributed vortices at two sides of the periodically nucleating bubbles were formed, as shown in **Fig. 5A**. The path lines in the fluid flow are constructed PS particle trajectories used to demonstrate the induced fluid flow by the periodically nucleated bubbles. An average flow speed of 14 mm/s was steadily achieved along the process in a selected area shown in **Fig. 5B**. This is about two orders of magnitude faster than the average flow speed previously reported(*26*).

In the second configuration, a glass sphere of 82 $\mu$m in diameter was fixed on the GNP decorated sample surface, as shown in **Fig. 5C**. This forms a minimal system for periodic bubble nucleation and can provide a localized point-and-shoot disturbance at the solid-liquid interface. When the periodic nucleation took place near the sphere, the liquid rapidly flowed from the bubble side toward the sphere side, leading to a unidirectional interfacial flow in several millimeter range. As shown in **Fig. 5D**, the measured average flow speed in the selected area is about 10 mm/s, which is also very strong compared with that achieved in current microfluidic applications. This bubble-sphere unit is more like a portable pump in microfluidics, which could be placed in a desired location in microchannels to implement pumping operations.

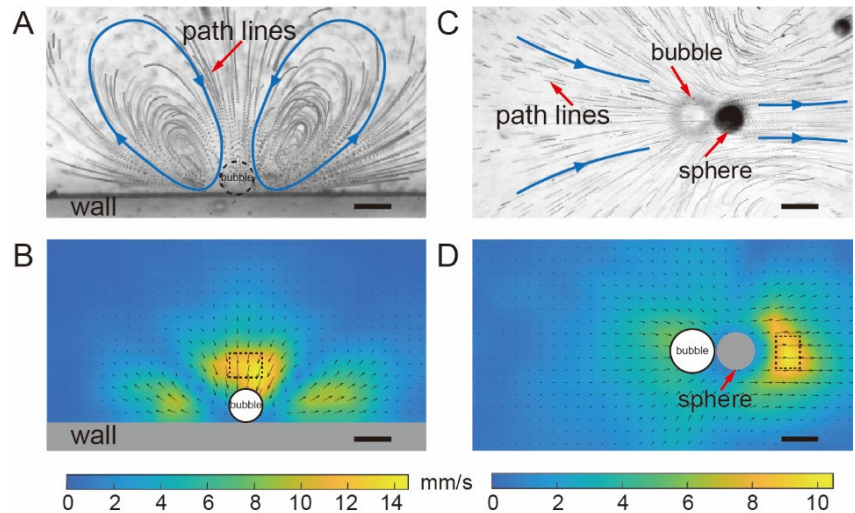

**Fig. 5** Strong localized flows at solid-liquid interfaces triggered by the periodically nucleated bubbles. (A) Strong interfacial vortex flows induced at the two sides of periodically nucleated bubbles near a vertical side wall captured from bottom view. (B) Constructed velocity field through the PIV measurement. The average flow speed along the periodic bubble nucleation in the selected area is about 14 mm/s. (C) A localized unidirectional interfacial flow induced by the periodically nucleated bubbles near a stationary sphere. (D) Velocity field obtained through the PIV measurement. The average fluid speed in the selected area is about 10 mm/s. (all scale bars: 100 $\mu$m)

In conclusion, we report an observation of spontaneously triggered periodic nucleation of plasmonic vapor bubbles on a gold nanoparticle decorated substrate. The presence of a nearby solid boundary results in the asymmetric collapse of plasmonic vapor bubbles, leading to the lateral locomotion of the residual bubbles. When the displacement is large enough to ensure the escape of the residual bubbles from a laser spot induced heat affected zone, consecutive bubble nucleation takes place. By adjusting laser power and laser spot size, the periodic bubble nucleation frequency $f$ can be readily modulated from 0.8 kHz to over 200 kHz, in a good agreement with the numerical prediction. Moreover, experimental results show that sustained localized fluid flows of over 10 mm/s can be steadily achieved by using the periodically nucleated microbubbles. This demonstrates how fast strong localized fluid flow can be achieved with photothermal bubbles and paves a way to achieve efficient micro/nanofluidic operations, using an ordinary continuous wave laser.

**Acknowledgments**

We thank Yue Chen for the artwork in the figures. We thank Buyun Chen for very helpful discussions of algorithm.

**Funding:** This work was supported by Natural Science Foundation of China (Grant No. 52075029), Beijing Natural Science Foundation (Grant No. 3232009) and "the Fundamental Research Funds for the Central Universities".

**Author contributions:** F. W., Y. W. and L. Y. conceived and designed the experiments; F. W., H. W. and C. X. performed the experiments; All the authors contributed to the analysis of the experimental data and discussed the results; Y. W., H. W. and F. W. wrote the manuscript; Y. W. and L. Y. supervised the project.

**Competing interests:** The authors declare no competing interests.

**Data and materials availability:** Data that support the plots within this paper and other findings of this study are available from the corresponding author upon reasonable request.

# Supplementary Materials for

# Sub-megahertz nucleation of plasmonic vapor microbubbles by asymmetric collapse

Fulong Wang[1], Huimin Wang[2], Binglin Zeng[1], Chenliang Xia[1], Lihua Dong[1], Lijun Yang[3,4], and Yuliang Wang[*,1,4]

[1]*Robotics Institute, School of Mechanical Engineering and Automation, Beihang University, Beijing 100191, People's Republic of China*

[2]*National Center for Materials Service Safety, University of Science and Technology Beijing, Beijing 100083, People's Republic of China*

[3]*School of Astronautics, Beihang University, Beijing 100191, People's Republic of China*

[4]*Ningbo Institute of Technology, Beihang University, Ningbo 315832, People's Republic of China*

[*]E-mail: wangyuliang@buaa.edu.cn

**This PDF file includes:**

    Supplementary Text
    Figs. S1 to S4
    References (1 to 6)

**Sample preparation**

A gold layer of ~12 nm in thickness was deposited on an amorphous fused silica substrate by a sputter coater (108 Auto, Cressington, UK). The gold film coated substrate was then put into a muffle furnace for thermal dewetting under an annealing temperature of 1000 ℃ for 1 hour. By applying this high temperature thermal dewetting, a gold nanoparticle (GNP) decorated sample surface was obtained, as illustrated in **Fig. S1A(I)**. **Fig. S1A(II)** shows a scanning electron microscopy (SEM) image of the obtained sample surface. The optical absorption spectra of the fabricated GNP decorated sample surface was measured by using a UV-Vis spectrophotometer (U-3900H, Hitachi, Japan). The result is depicted in **Fig. S1A(III)**. A peak absorption value of 0.332 was achieved at the wavelength of 529 nm, which is very close to that of 532 nm of the applied laser.

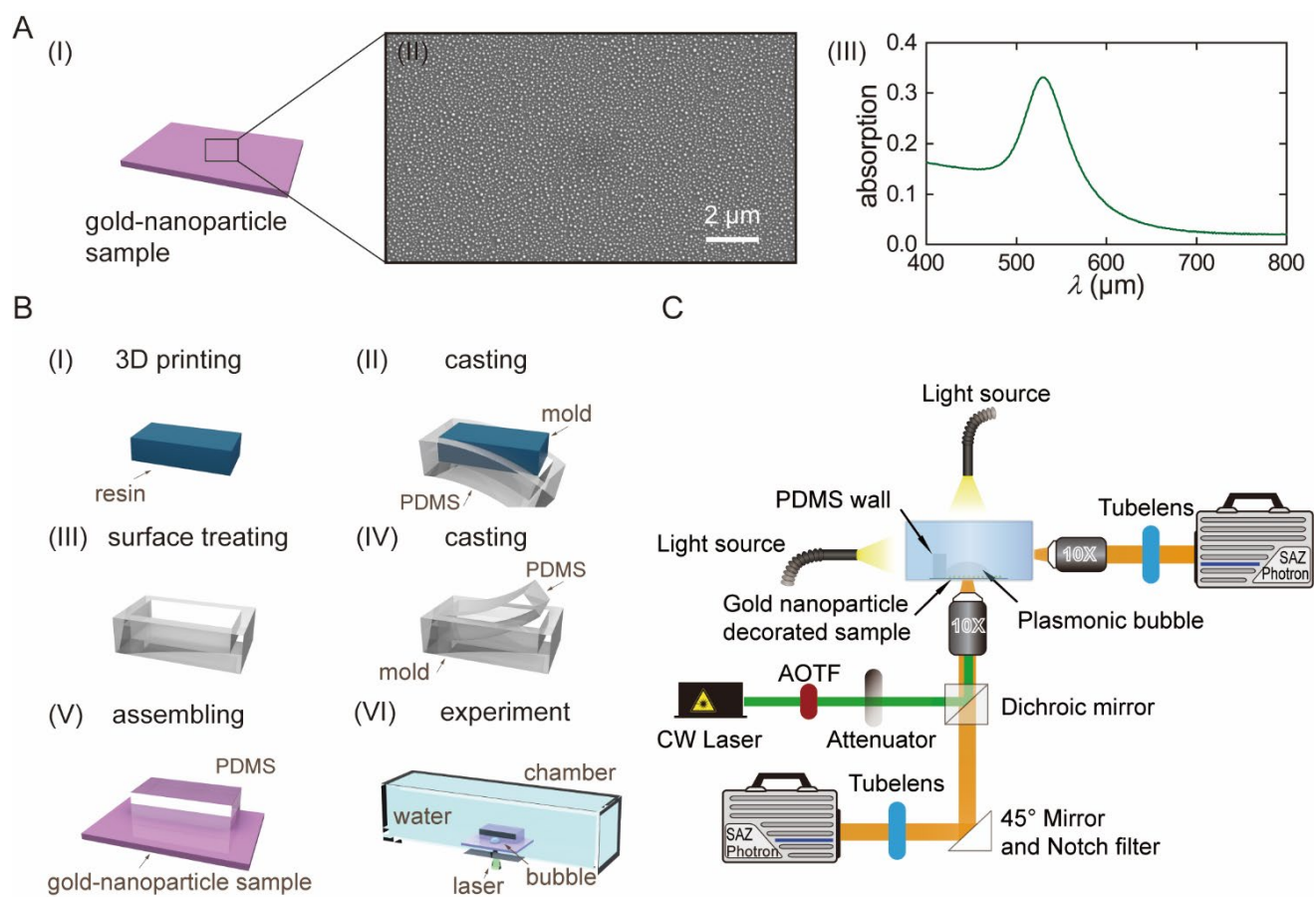

**Fig. S1** Sample preparation and optical setup for vapor bubble experiments. (A, I) Schematic of a gold-nanoparticle (GNP) decorated sample surface and (A, II) its SEM image. (A, III) Experimentally measured optical absorption spectrum of the GNP decorated sample surface. (B) Schematic diagram of the fabrication process of a vertical side wall on the GNP decorated sample surface. (C) Schematic diagram of the optical setup for vapor bubble experiments.

The process to fabricate the polydimethylsiloxane (PDMS) vertical side wall on the GNP

decorated sample surface is illustrated in **Fig. S1B**. In the first step, a male mold was fabricated *via* a high-resolution 3D printing of a photosensitive resin (**Fig. S1B(I)**). In the second step, a female mold was obtained by taking the male mold as the pattern in a mixture of the Sylgard 184 silicone elastomer base agent and its curing agent (Dow Corning, USA) at 65°C for 3 hours (**Fig. S1B(II)**). The female mold was then coated with WD-40 to facilitate the subsequent mold separation (**Fig. S1B(III)**). In the third step, a PDMS bar was fabricated by pouring a mixture of PDMS and its curing agent into the female mold obtained in the second step (**Fig. S1B(IV)**). By gently pressing, the PDMS bar was firmly attached to the GNP decorated sample surface, providing a vertical side wall (**Fig. S1B(V)**). The substrate was then put into a water-filled cuvette with a dimension of 10×10×75 mm$^3$ for vapor bubble experiments (**Fig. S1B(VI)**).

**Setup description**

A schematic diagram of the developed optical setup for bubble experiments is shown in **Fig. S1C**. In the setup, two high-speed cameras (SA-Z, Photron, Japan) were used for simultaneous side-view and bottom-view imaging. Long working distance objective lenses with different magnifications of 5×, 10×, and 20× (LMPFLN, Olympus, Japan) were equipped to adjust laser spot size on the sample surface. The frame rate of high speed imaging is 900 kfps. A continuous wave laser was applied to trigger nucleation of plasmonic bubbles. The laser power was tuned *via* a half-wave plate and a polarizer. An acousto-optic modulator (AOTFnc-VIS-Tn, Opto-Electronic, France) was used as a shutter to control on/off of the laser irradiation on the sample surface.

**Determination of cycle period *T* of the periodic bubble nucleation**

The temporal resolution of the optical imaging system is 1.11 μs. The finite temporal resolution may introduce a relatively large error in determining cycle period *T* of the periodic bubble nucleation, especially when *T* is below 10 μs. To solve this problem, the measured bubble radius *R* as a function of time *t* is fitted to an analytical approximation of the Rayleigh equation (*1*)

$$R(t)/R_{max} = \left(1 - \left(t/\frac{\tau_c}{2}\right)^2\right)^{2/5} \tag{S1}$$

Through this fitting, the maximum bubble radius $R_{max}$, as well as corresponding moment $t_{max}$ at which the bubble reaches $R_{max}$, can be preciously obtained. The period of time between the two neighboring values of $t_{max}$ is then taken as cycle period *T*, as demonstrated for two successively nucleated vapor bubbles in **Fig. S2**. With the obtained *T*, the nucleation frequency *f* can be further determined as *f* = 1/*T*. By using this method, the extracted *f* in **Fig. 4C** has a very low variance even when the cycle period *T* is below 10 μs.

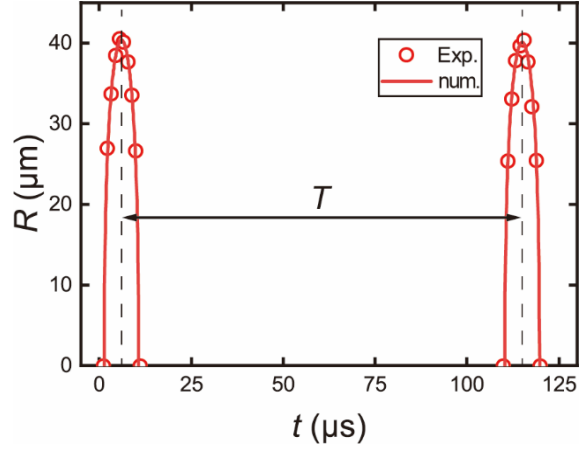

**Fig. S2** The radius of two successively nucleated bubbles as a function of time *t*. The hollow circular markers are experimental data points of bubble radius. The data points were fitted to an analytical approximation of the Rayleigh equation (Eq. S1) and the fitting results are drawn by the two solid curves. From the obtained fitting curves, the maximum bubble radius $R_{max}$ as well as corresponding moment $t_{max}$ can be extracted for individual vapor bubbles. The period of time in between two neighboring moments of $t_{max}$ can then be taken as cycle period *T*.

**Determination of laser spot radius**

The determination of laser spot radii on the GNP decorated sample surface is important for the numerical investigation of temperature fields under laser irradiation. To do so, the captured laser spot images on the GNP decorated sample surface were fitted to a 2D Gaussian function. The colored solid markers in **Fig. S3** represent the normalized light intensity of an experimentally captured laser spot image from the 10× objective lens. The transparent mesh surface represents the fitting result. After that, the domain with light intensity over $1/e^2$ of the maximum value is taken as the laser spot area, where *e* is the natural exponent. By using this method, the obtained radii of laser spots are about 11.5 $\mu$m and 28.0 $\mu$m for the 10× and 5× objective lenses, respectively.

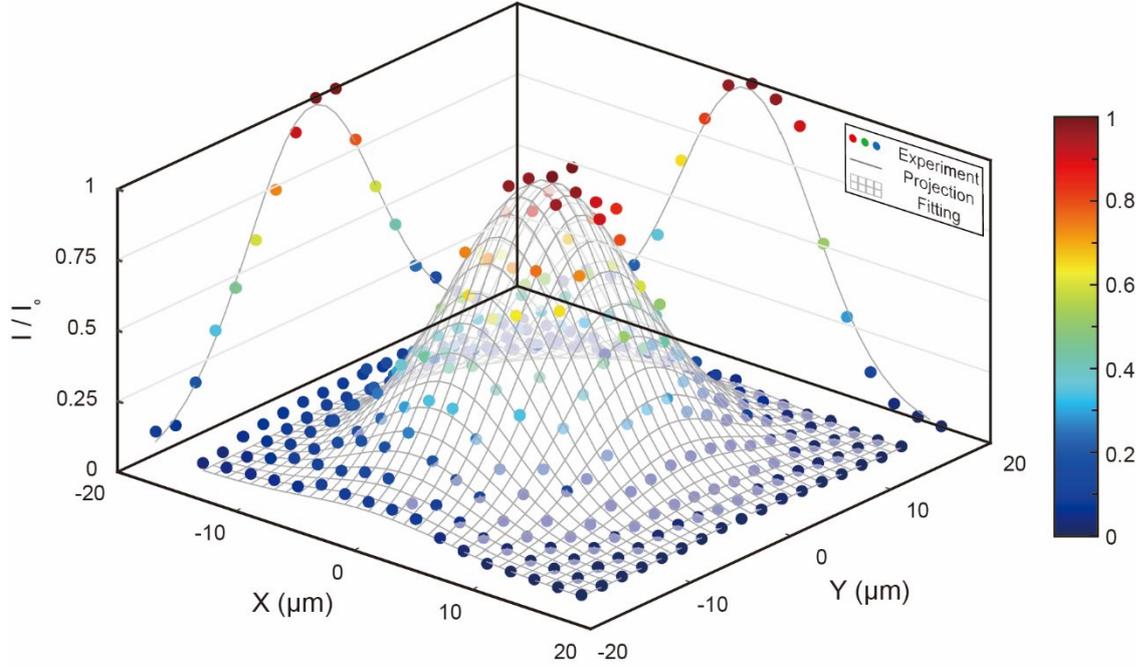

**Fig. S3** Fitting of the normalized light intensity in a laser spot image of the 10× objective lens to a 2D Gaussian function. The obtained 2D Gaussian fitting surface agrees well with the experimental data. From the fitting, laser spot radii were determined.

**The BIM simulation of bubble dynamics**

Here we briefly describe the BIM simulation. Please refer to Blake et al (*2, 3*) and Supponen et al (*4*) for detailed description. The BIM simulation is based on a standard model of an inviscid, incompressible and irrotational flow with negligible surface tension. It is governed by mass conservation and the simplified Navier-Stokes equation, given as:

$$\nabla \cdot \boldsymbol{u} = 0 \tag{S2}$$

$$\frac{\partial \boldsymbol{u}}{\partial t} + (\boldsymbol{u} \cdot \nabla)\boldsymbol{u} = -\frac{\nabla p}{\rho} + \boldsymbol{g} \tag{S3}$$

where $\boldsymbol{u}$ is the velocity field of the fluid, $p$ the pressure, $\boldsymbol{g}$ the gravitational acceleration, $\rho$ the density of water.

Due to the assumption of irrotational flow, i.e., $\nabla \times \boldsymbol{u} = 0$, the velocity field $\boldsymbol{u}$ of the fluid can be derived from the derivative of potential $\varphi$

$$\boldsymbol{u} = \nabla \varphi \tag{S4}$$

According to Eqs. S2 and S4, the fluid potential $\varphi$ satisfies the Laplacian equation:

$$\nabla^2 \varphi = 0 \tag{S5}$$

For any sufficiently smooth function $\varphi$ which satisfies Laplacian equation in a domain $\Omega$ having piecewise smooth surface $S$, it can be rewritten as follows by applying Green's integral formula:

$$\lambda \varphi(P) = \iint_S \left( \frac{\partial \varphi(Q)}{\partial n} \frac{1}{|P-Q|} - \varphi(Q) \frac{\partial}{\partial n}\left(\frac{1}{|P-Q|}\right) \right) dS \tag{S6}$$

where $P \in \Omega + S$ and $Q \in S$ are the reference and the integral points, respectively, $\lambda$ is the observation spatial angle from point $P$ and $\partial/\partial n$ is the outward normal derivative of $S$. The angle $\lambda$ varies with the position of point $P$ and is given as

$$\lambda = \begin{cases} 2\pi & P \in S \\ 4\pi & P \in \Omega \end{cases} \tag{S7}$$

If both points are known on $S$, Eq. S6 can be solved to get $\varphi$ at any point $P$. So, combined with Eq. S4, the velocity at any position in the fluid can be obtained from the boundary potential $\varphi$ and its normal derivative $\partial \varphi / \partial n$.

The temporal evolution of the potential $\varphi$ can be given based on Bernoulli's principle

$$\frac{D\varphi}{Dt} = \frac{|u|^2}{2} - gz + \frac{\Delta p}{\rho} \tag{S8}$$

where $D/Dt$ is the material derivative. By numerically solving Eqs. S6 and S8, the shape of bubbles and the potential $\varphi$ on its surface can be updated for every time step in the simulation space.

**Numerical solution of temperature field around laser spots**

The temporal evolution of water temperature field in the vicinity of the laser spot plays a key role in governing bubble nucleation and growth dynamics. The time-dependent temperature field $T(r, t)$ of the surrounding liquid caused by a single nanoparticle can be obtained by numerically solving the following spherical linear Fourier equation of heat conduction:

$$\partial_t (T(r,t)) = \frac{p_l(r,t)}{\rho c_p} + \kappa \frac{1}{r^2} \partial_r \left( r^2 \partial_r T(r,t) \right) \tag{S9}$$

where $\kappa$ and $c_p$ are thermal diffusivity and heat capacity of water, $r$ is the spherical distance to the GNP, and $p_l(r, t)$ is the deposited power density (unit in W / m³), which is assumed to be constant for a radius $r$ within the GNP, and 0 elsewhere. The linear superposition of the temperature fields for individual GNPs within the Gaussian laser beam profile can then be taken as the temperature field around the laser spot area, given as

$$T(x,y,z,t) = \sum_{i=1}^{N_{np}} \left[ T_i \left( d_{i,(x,y,z)}, t \right) \right] \tag{S10}$$

where $N_{np}$ is the number of GNPs under laser irradiation, $T_i$ is the temperature field produced by the $i_{th}$ nanoparticle, and $d_{i,(x,y,z)}$ is the distance from the center of the $i_{th}$ nanoparticle to the point located at the coordinates $(x, y, z)$. By using this model, the spatial-temporal evolution of the liquid temperature in the vicinity of the laser spot for any specific laser power $P_l$ was obtained.

The peak temperature as a function of time $t$ under different laser power $P_l$ for the 10× objective lens is given in **Fig. S4A**. For a given laser power $P_l$, one can directly obtain the time required to reach a specific temperature from the room temperature $T_E = 20$ °C (for example, the upper horizon line in **Fig. S4A**). The experimentally measured delay $\tau_d$ in **Fig. 4A** was fitted to the constructed $\Delta T - t$ dependence in **Fig. S4A** to estimate the nucleation temperature $T_n$ by using the root-mean-square-minimization method. Through this approach, the estimated nucleation temperature $T_n$ is around 215 °C. When the maximum temperature in the laser spot area reaches $T_n$, an IPB will be nucleated.

During periodic bubble nucleation, the sustained laser energy input leads to an overall temperature elevation of the fluid field compared to the ambient temperature $T_E$ in the vicinity of the laser spots. This overall temperature elevation will inevitably change the nucleation dynamics of a periodically nucleated bubble compared to the first triggered IPB. A major difference is the delay $\tau_d$. Since the nucleation temperature $T_n$ is constant, this overall temperature elevation definitely leads to decreased $\tau_d$ for periodically nucleated IPBs. Here we estimate this overall temperature elevation. This helps us to evaluate how $f$ changes with $P_l$ (**Fig. 4C**) and then to define the regimes of repeatable bubble nucleation in **Fig. 4D**.

To assess this temperature elevation, the delay $\tau_d$ of the periodically nucleated bubbles was first measured. As mentioned in the main text, the cycle period $T$ contains two periods, a delay time $\tau_d$ and the life time $\tau_c$ of a vapor bubble. The cycle period $T$ and life time $\tau_c$ were measured by the method described above. By subtracting $\tau_c$ from $T$, the delay time $\tau_d$ at different $P_l$ was obtained. With the measured $\tau_d$, the elevated ambient temperature $T'_E$ was obtained for each $P_l$, assuming the constant nucleation temperature $T_n$. Then the overall temperature elevation can be given as $\Delta T_E = T'_E - T_E$. We find that different $P_l$ lead to different temperature elevation $\Delta T_E$. Surprisingly, the result show that $\Delta T_E$ linearly increases with $P_l$ (**Fig. S4A inset**). **Fig. S4B** demonstrates a typical temperature rise process. The nucleation center is heated up from the elevated water temperature of $T_E + \Delta T_E$ to the nucleation temperature $T_n$ in a laser irradiation period of $\tau_d$.

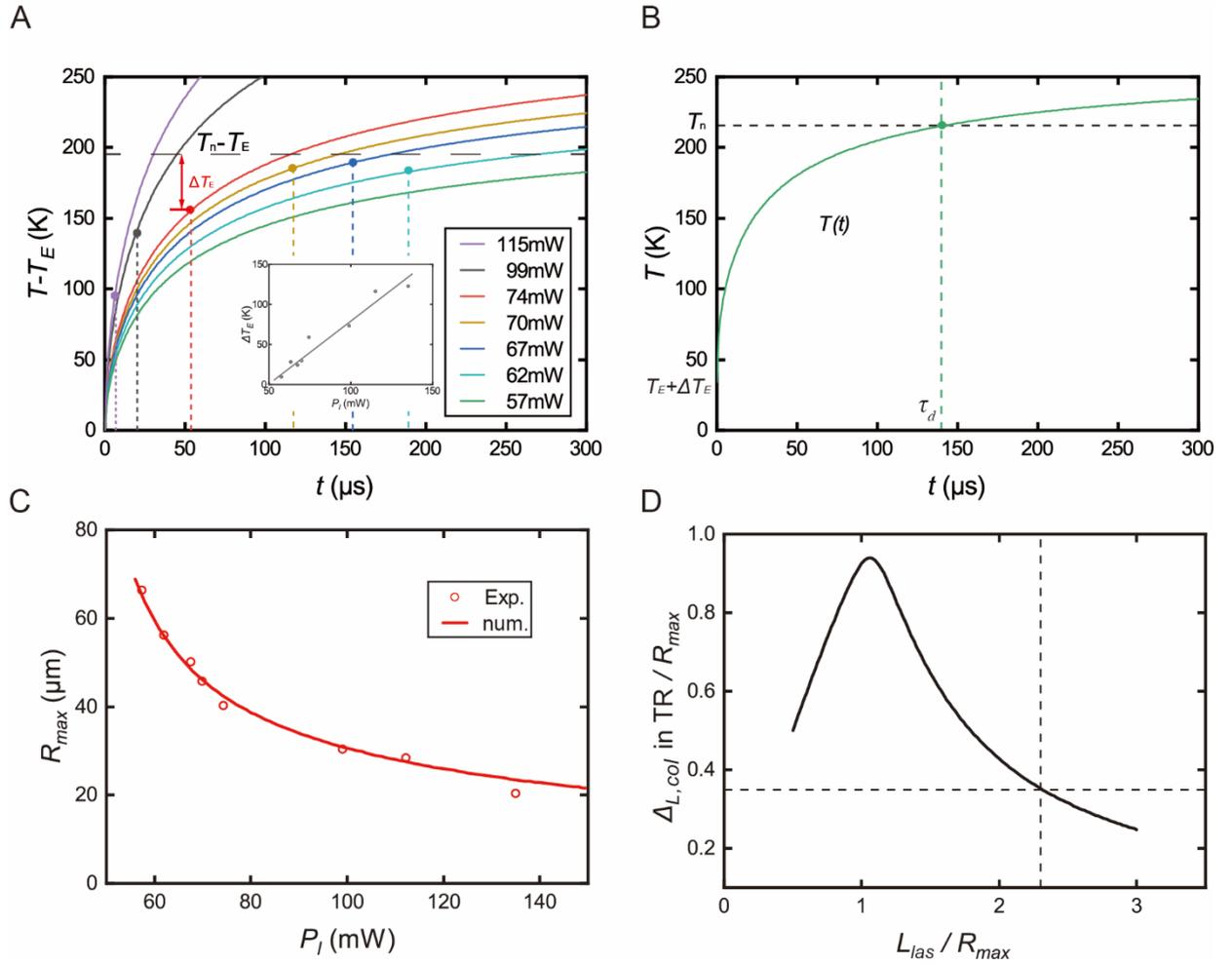

**Fig. S4** (A) Temperature rise at the center of the laser spot as a function of time $t$ under different laser power $P_l$. The inset figure shows how $\Delta T_E$ changes with $P_l$. (B) A typical temperature rise curve at the center of the laser spot as a function of $t$. (C) The maximum size $R_{max}$ as a function of $P_l$ from experimental measurement (markers) and numerical calculation (solid curve). (D) $\Delta_{L,col}/R_{max}$ as a function of $L_{las}/R_{max}$ for IPBs constructed by the numerical simulation.

With the estimated $\Delta T_E$, bubble size $R_{max}$ under any specific laser power $P_l$ can be numerically determined, which is necessary for the theoretical prediction of bubble lifetime $\tau_c$ and hence the nucleation frequency $f$ shown in **Fig. 4C**. To do so, we applied a method reported in the publication (*5*) and chose the vaporization temperature $T_{vap}$ = 141°C at 1 bar. For each laser power $P_l$, a temperature field was first constructed using the above mentioned method. Then the amount of water molecules of which the temperature is over $T_{vap}$ was determined from the constructed temperature field. After that, the bubble volume $V_{max}$ as well as its maximum radius $R_{max}$ can be determined by using the ideal gas law. The solid line in **Fig. S4C** is the numerically determined $R_{max} - P_l$ dependence obtained for the 10× objective lens. It clearly shows that the numerical simulation results agree well with our experimentally measured $R_{max}$. After $R_{max}$ was obtained, the bubble lifetime $\tau_c$ was further obtained by

using a modified Rayleigh collapse time formula (*6*) and by taking a prolongation factor *k* = 1.3, given as

$$\tau_c = k \cdot 2t_c = 1.3 \times 2 \times 0.915 \times R_{max}\sqrt{\rho/(P_0 - P_v)} \tag{S11}$$

where $P_0$ and $P_v$ are the ambient pressure and the saturated vapor pressure, respectively. With the numerically determined $\tau_d$ and $\tau_c$, the theoretical nucleation frequency *f* can be obtained, as shown in **Fig. 4C**.

Both bubble maximum radius $R_{max}$ and the critical values of $\overline{\Delta}_{L,col}$ in the transition region were used to theoretically determine the available regimes for repeatable vapor bubble nucleation in the $P_l$ - $L_{las}$ parameter space (**Fig. 4D**). For a specific objective lens, $R_{max}$ at a certain $P_l$ can be determined by using the method mentioned above. Then the ratio of $\overline{\Delta}_{L,col}$ in transition region (**Fig. 3B**) to $R_{max}$ was used to determine corresponding threshold value of $L_{las}/R_{max}$ in **Fig. S4D**. Accordingly, the threshold value of $L_{las}$ can be determined. Below this threshold value, IPBs are supposed to be periodically nucleated. By using this method, the boundaries of repeatable nucleation regimes in the $P_l$ - $L_{las}$ parameter space were determined (**Fig. 4D**).

**PIV measurement**

For the PIV measurement in **Fig. 5** of main text, polystyrene (PS) particles (2 *μ*m in diameter) were added into deionized (DI) water with a concentration of 60 *μ*g/mL. The SA-Z high-speed camera equipped with the 10x objective lens was used to record the images at a frame rate of 20 kfps during the PIV measurements. These images were processed by PIVlab toolbox in MATLAB to construct the flow field. Square interrogation windows of 32×32 pixels with an overlap of 50% were used to obtain the velocity vectors.